# Circulating tumor cell detection in cancer patients using in-flow deep learning holography

Kevin Mallery[1,★], Nathaniel R. Bristow[1,★], Nicholas Heller[1,2], Yash Travadi[1,3], Ali Arafa[4], Kaylee Kamalanathan[1], Catalina Galeano-Garces[1], Mahdi Ahmadi[1], Grant Schaap[1], Alexa Hesch[1], Olivia Hedeen[1], Zikora Izuora[1], Joel Hapke[1], Jeffrey Miller[1], Arjun Viswanathan[1], Ivo Babris[1], Songyi Bae[1], Tuan Le[1], Tony Clacko[1], Emmanuel S. Antonarakis[5,6], Badrinath R. Konety[1,7], Justin M. Drake[1,4,8], Jiarong Hong[1,9,10,†]

[1]Astrin Biosciences, St Paul, MN, USA
[2]Department of Computer Science and Engineering, College of Science and Engineering, University of Minnesota – Twin Cities, Minneapolis, MN, USA
[3]School of Statistics, College of Liberal Arts, University of Minnesota – Twin Cities, Minneapolis, MN, USA
[4]Department of Pharmacology, Medical School, University of Minnesota – Twin Cities, Minneapolis, MN, USA
[5]Department of Medicine, Medical School, University of Minnesota – Twin Cities, Minneapolis, MN, USA
[6]Masonic Cancer Center, University of Minnesota – Twin Cities, Minneapolis, MN, USA
[7]Allina Health Cancer Institute, Minneapolis, MN, USA
[8]Department of Urology, Medical School, University of Minnesota – Twin Cities, Minneapolis, MN, USA
[9]Department of Mechanical Engineering, College of Science and Engineering, University of Minnesota – Twin Cities, Minneapolis, MN, USA
[10]Department of Electrical and Computer Engineering, College of Science and Engineering, University of Minnesota – Twin Cities, Minneapolis, MN, USA

★These authors contributed equally
†Corresponding author: jiarong.hong@astrinbio.com and jhong@umn.edu

## Abstract

Circulating tumor cells (CTCs) are cancer cells found in the bloodstream that serve as biomarkers for early cancer detection, prognostication, and disease monitoring. However, CTC detection remains challenging due to low cell abundance and heterogeneity. Digital holographic microscopy (DHM) offers a promising, label-free method for high-throughput CTC identification by capturing superior morphological information compared to traditional imaging methods, while remaining compatible with in-flow data acquisition. We present a streamlined DHM-based system that integrates microfluidic enrichment with deep learning-driven image analysis, supplemented by immunofluorescent profiling, to improve sensitivity and specificity of CTC enumeration. Specifically, our platform combines inertial microfluidic preprocessing with dual-modality imaging, integrating holography with fluorescence sensing of up to two markers. A deep learning model, trained on a diverse set of healthy blood samples and cancer cell lines, and executed in real-time, provides a morphological confidence on a cell-by-cell basis that may then be combined with immunofluorescence criteria for enumeration. In a pilot study, we demonstrate significantly higher CTC counts in patients with late-stage prostate cancer (n=13) compared to healthy controls (n=8), with a patient-level false positive rate of 1 cell/mL. Notably, nearly two-thirds of identified CTCs were EpCAM-negative but PSMA positive (a prostate specific epithelial marker), suggesting that traditional use of EpCAM as an epithelial marker for CTCs may lead to false negatives. These findings highlight the potential of DHM for applications including but not limited to screening, diagnostics, and precision oncology.

## Introduction

Circulating tumor cells (CTCs) are malignant cells that detach from primary or metastatic tumors and enter the bloodstream, where they act as precursors to distant metastasis and serve as real-time indicators of disease status (Alix-Panabières et al. 2014a, Massague et al. 2016). As early as stage 0, CTCs can be detected in peripheral blood, reflecting the potential for dissemination before radiographic or symptomatic manifestation (Bae et al. 2024). Their abundance and phenotypic characteristics have been linked to tumor aggressiveness, therapeutic resistance, and patient prognosis (Cristofanilli et al. 2004, De Bono et al. 2008), making them a valuable biomarker for early cancer detection, risk stratification, and longitudinal monitoring. Unlike tissue biopsies, CTCs can be accessed noninvasively and repeatedly, enabling dynamic insight into tumor evolution and therapeutic response over the course of treatment (Pantel et al. 2019).

However, detecting and enumerating CTCs via cytometry, biomarker labelling, or molecular analysis remains challenging due to their extreme rarity and phenotypic variability (Alix-Panabières et al. 2014a). Specifically, a single milliliter of blood contains approximately $10^6$ white blood cells (WBCs) and $10^9$ red blood cells (RBCs) (Tefferi et al. 2005, Cleveland Clinic 2024), but may contain fewer than one CTC in early-stage cancers (Pantel et al. 2010, Yu et al. 2013) and rarely more than ten CTCs even in late-stage patients (Cristofanilli et al. 2004). CTC surface marker expression is highly heterogeneous. Fluorescent markers such as EpCAM and cytokeratin may identify only a subset of tumor cells, particularly those undergoing epithelial-to-mesenchymal transition (EMT), during which epithelial markers are frequently downregulated (Mikolajczyk et al. 2011, Yu et al. 2013). Beyond molecular variability, CTCs also differ widely in physical properties, including size, deformability, and density. These features can substantially overlap with those of WBCs, further complicating accurate discernment (Gascoyne et al. 2009, Xu et al. 2017, Khoo et al. 2020).

Despite the rarity and phenotypic heterogeneity of CTCs, numerous detection methods have been developed that rely on molecular markers or physical traits, broadly categorized as antigen-based and biophysical separation approaches (Alix-Panabières et al. 2014b, Ferreira et al. 2016, Shen et al. 2017, Ju et al. 2022). Among antigen-based methods, CellSearch® remains the most clinically validated platform, enriching EpCAM-positive cells via immunomagnetic capture (Allard et al. 2004), with prognostic utility established in metastatic cancers (Cristofanilli et al. 2004, de Bono et al. 2008). Subsequent microfluidic platforms like the CTC-Chip (Nagrath et al. 2007) and Herringbone-Chip (Stott et al. 2010) improved sensitivity through micropost arrays and passive mixing. Recent advancements include aptamer-functionalized nanostructures (Ding et al. 2020), droplet microfluidic single-cell analysis (Chen et al. 2022), and 3D micro-/nanostructured interfaces that enhance capture efficiency via fluid dynamics (Kang et al. 2022). However, antigen-based methods often fail to capture phenotypically heterogeneous CTC subpopulations (Mikolajczyk et al. 2011, Yu et al. 2013), while microfluidic platforms with complex architectures face challenges in large-scale implementation (Kang et al. 2022).

Various biophysical methods have been developed to isolate CTCs by exploiting intrinsic properties such as size, density, deformability, and electrical or mechanical characteristics (Harouaka et al. 2013, Ju et al. 2022). Specific techniques include size-based filtration such as the early ISET platform (Vona et al. 2000), inertial microfluidics exemplified by the FDA-cleared Parsortix® system (Miller et al., 2018), and deterministic lateral displacement (DLD) (Okano et al. 2015, Liu et al. 2021), as well as dielectrophoresis (Gascoyne et al. 2009) and acoustic separation (Li et al. 2015). However, the wide variability in CTC phenotypes and their physical overlap with WBCs often leads to limited specificity, reduced purity, and potential loss of CTCs in clinical implementation (Harouaka et al. 2013, Habli et al. 2020, Ju et al. 2022).

Hybrid strategies have emerged to address the limitations of single-modality CTC detection by combining biophysical and immunoaffinity-based mechanisms (Deng et al. 2022). Examples include the CTC-iChip integrating inertial focusing with magnetic WBC depletion (Ozkumur et al. 2013), size-dictated immunocapture chips enabling geometric enrichment and profiling (Ahmed et al. 2017), a high-throughput inertial–magnetic sorter for leukapheresis-scale processing (Mishra et al. 2020), and integrated inertial–magnetophoretic microdevices (Nasiri et al. 2022). While these systems improve sensitivity and subtype coverage, they remain limited by device complexity, high cost, and cumulative CTC loss. These challenges highlight the need for streamlined, high-fidelity platforms compatible with clinical workflows.

Digital holographic microscopy (DHM) has emerged as a promising alternative for label-free, high-content imaging of biological samples (Xu et al. 2001). As a coherent imaging technique, DHM captures both the amplitude and phase of transmitted light by recording its interference with a reference wave (Tahara et al. 2018). Compared to bright-field and other incoherent imaging methods, DHM provides a significantly larger depth of focus and encodes optical thickness and 3D spatial data, supporting applications such as cell culture monitoring, microbe classification, and particle tracking (Kastl et al. 2017, Göröcs et al. 2018, Kumar et al. 2025). Despite its advantages, recent explorations of DHM for CTC detection remain constrained by limited throughput, poor specificity at the patient level, and a lack of validation using clinical blood samples (Pirone et al. 2023; Gangadhar et al. 2023).

To address these challenges, we developed a clinically scalable platform that integrates real-time DHM with inertial microfluidic enrichment and orthogonal immunophenotyping. The system is compact, cost-effective, and readily parallelizable, enabling rapid processing of standard 10 mL blood tubes within hours. A customized deep learning model trained on millions of healthy blood cells and a diverse panel of cancer cell lines enables robust classification of patient-derived images in flow. The platform achieves a low patient-level false positive rate of 1 cell/mL and supports downstream analyses such as single-cell multi-omics and 3D tissue culture. In the first large-cohort application of DHM for CTC detection, our method distinguishes late-stage prostate cancer patients from healthy controls, demonstrating its potential as a clinically viable tool for cancer diagnostics and screening.

# Results

## Integrated platform for high-precision, high-throughput in-flow CTC detection

Our platform unites three orthogonal technologies—patented saw-tooth inertial microfluidics, DHM, and dual-channel immunofluorescence (IF)—into an optimized in-flow process that overcomes both the low abundance and phenotypic heterogeneity of CTCs (Fig. 1). To begin, whole blood is diluted in a proprietary buffer media aimed at minimizing cell loss through adhesion to plastic surfaces. Next, it flows through Astrin's patented saw-tooth inertial microfluidic chip (WO 2024/064911 A1), which enriches CTCs by depleting hematologic background. This process removes over 99.999% of RBCs and approximately 99.6% of WBCs (Fig 2), while retaining 95% of spiked cancer cells (Supplementary Information Section 1, SI §1). Overall, this results in more than a 100-fold enrichment of CTCs in the processed sample (Fig. 2). A detailed description of the microfluidic chip and sample preparation protocols is provided in Methods.

Subsequently, the enriched sample is hydrodynamically focused into a straight microfluidic channel and analyzed using our dual-modality imaging system (details in Methods). A digital camera captures the holograms generated by a pulsed 405 nm laser, encoding both the optical and morphological information of the cells (Fig. 1). A customized deep learning neural network processes these holograms in real-time (as described in the next section and detailed further in Methods). The network produces a spatial probability heatmap highlighting potential CTC locations, as learned from training on annotated images of cancerous and healthy cells.

Such in-flow CTC classification via holographic imaging provides an avenue for a completely label-free enrichment platform and can improve over time as training datasets grow and deep learning models evolve. Crucially, the label-free DHM system remains compatible with IF-based molecular labeling to enhance the accuracy of the system or provide secondary enumeration metrics (e.g., degree of EpCAM expression). For the system presented herein, two photomultiplier tubes (PMTs) are integrated to capture IF signals emitted by tumor-specific antibody conjugates upon excitation by a co-linear 488 nm laser (Fig. 1). A custom signal processing algorithm (SI §2.1) identifies true cellular fluorescence peaks amid background noise in each PMT channel, performs cross-channel matching, and integrates the resulting IF readouts with predictions from the DHM classifier.

Overall, the logical conjunction of biophysical profiling from deep learning-based DHM and biochemical readout from IF enables a dramatic reduction in false positives (Fig. 2). The DHM classifier and IF channel each independently provide 100-fold reduction of nucleated cells, yielding a total false positive rate on the order of 1 cell mL$^{-1}$ that is sufficient for cancer detection. Importantly, the dual-modality sensing design mitigates overreliance on antigen expression issues, such as EpCAM downregulation during EMT. CTCs that escape detection by antigen-based markers remain identifiable through their distinct holographic signatures, while spurious fluorescence signals arising from nonspecific antibody binding are filtered out by the deep learning model.

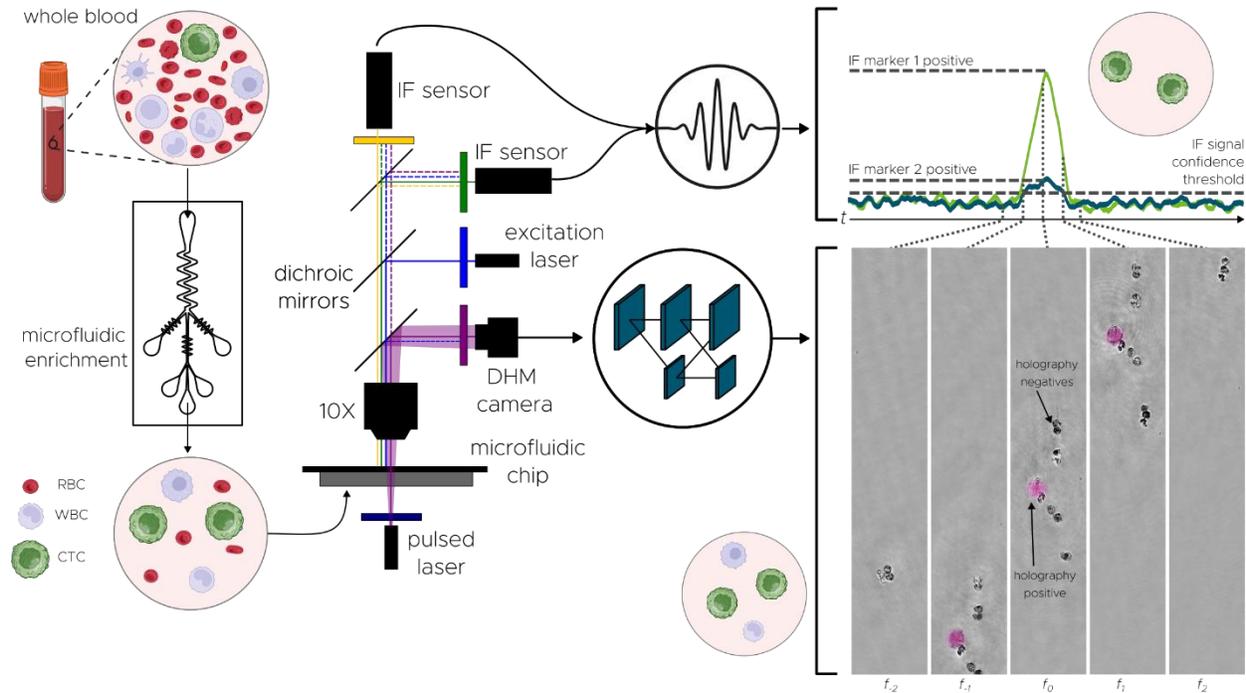

Fig. 1 System overview, depicting the detection of CTC candidates based on a combination of microfluidic inertial enrichment, digital holographic microscopy signature, and immunofluorescence (IF) expression. Samples begin as whole blood prior to microfluidic enrichment, during which red blood cells are primarily depleted while retaining CTCs and most WBCs. Subsequently, holograms of each cell and IF signal from the field-of-view as a whole are captured during passage through a second microfluidic chip. Detections in the IF signal for emitting cells are visible as broad peaks whose time-scale is inversely proportional to the frame rate (i.e., equivalent to the time taken for a cell to passage the field-of-view). Holograms pass through a neural network to classify each cell and detect CTCs, while the IF data can be used for further filtering operations for cell enumeration.

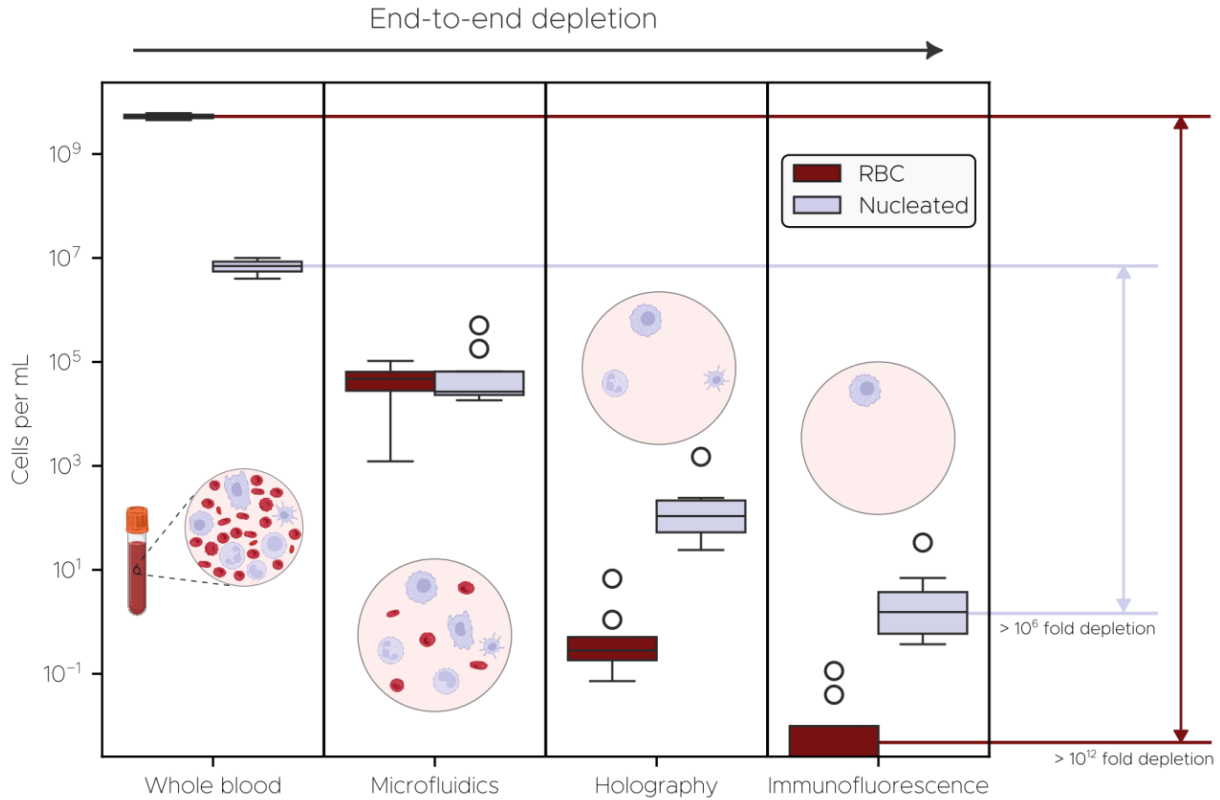

*Fig 2. Cumulative end-to-end system depletion performance. Cell counts for whole blood are based on Cleveland Clinic complete blood count data for red blood cells and white blood cells in males (Cleveland Clinic 2024). Microfluidic, holographic, and immunofluorescence depletion data are measured on the imaging system. Fold depletion is calculated in terms of means, rather than medians, as the median value for immunofluorescence stage is zero for RBCs. Each of the four stages shown here mirrors those depicted in Fig. 1.*

## Deep Learning Framework

The benefits of using DHM for label-free CTC detection are counterbalanced by several intrinsic challenges. Chief among these is the extreme scarcity of CTCs, which makes it difficult to compile a sufficiently large and diverse set of positive examples for training a robust detection model. Available public datasets are limited and primarily consist of cultured cancer cell lines. However, patient-derived CTCs exhibit far broader phenotypic heterogeneity than any single cell line can represent (Park et al. 2014, van der Toom et al. 2016). Figure 3 summarizes the strategy we adopted to address these challenges while maintaining the real-time throughput required for in-flow assays.

We formulated the detection task as an image-to-heatmap transformation in which a streamlined High-Resolution Network (HR-Net) (Wang et al. 2020, SI §3.1) directly yields a probability surface. Local maxima above a chosen confidence threshold correspond to putative cells. Representing cells as Gaussian key points—rather than bounding boxes— obviates the need for precise manual annotation of poorly-defined holographic fringes. The number and depth of model layers were optimized to strike a balance between descriptive depth and inference speed—ensuring both high prediction accuracy while supporting live processing of full-field holograms (architectural details in Methods).

Training data were deliberately assembled to counter both the rarity of CTCs and their morphological diversity. Two complementary streams were employed (Fig. 3, left side). The negative stream comprised enriched healthy blood, exposing the network to the full spectrum of WBCs, RBCs, debris, and background variation. The positive stream consisted of suspensions of five histologically distinct cancer cell lines imaged in buffer. Every nucleated object in these frames is a putative cancer cell. Because such images lack pixel-level labels, a companion HR-Net was first trained on select hand-annotated frames to create provisional "pseudo-labels". During subsequent detector training, images that produced misclassifications were resurfaced in successive epochs by hard-sample mining (further details in Methods), ensuring repeated exposure to the most confounding features.

An asymmetric cross-entropy loss embodied the biological class imbalance: weights were skewed towards healthy-blood examples to penalize WBC misclassification while remaining tolerant of occasional pseudo-labelling errors in the positive stream. Further details can be found Methods. After training, additional cancer lines—25 in total—were sequestered for validation, demonstrating that the network learns invariant features that generalize beyond the limited set of cultures available for optimization. The full list of cell lines, hyper-parameters, and augmentation protocols are provided in Methods and SI §4 Tables 2–3.

Together, the key-point formulation, HR-Net customization, heterogeneity-centered sampling, and asymmetric loss establish a conceptually distinct framework (Fig. 3, center and right panels) that narrows the gap between idealized cell-line imagery and the complex reality of patient-derived CTCs while sustaining the high frame rates required for population-scale liquid biopsy.

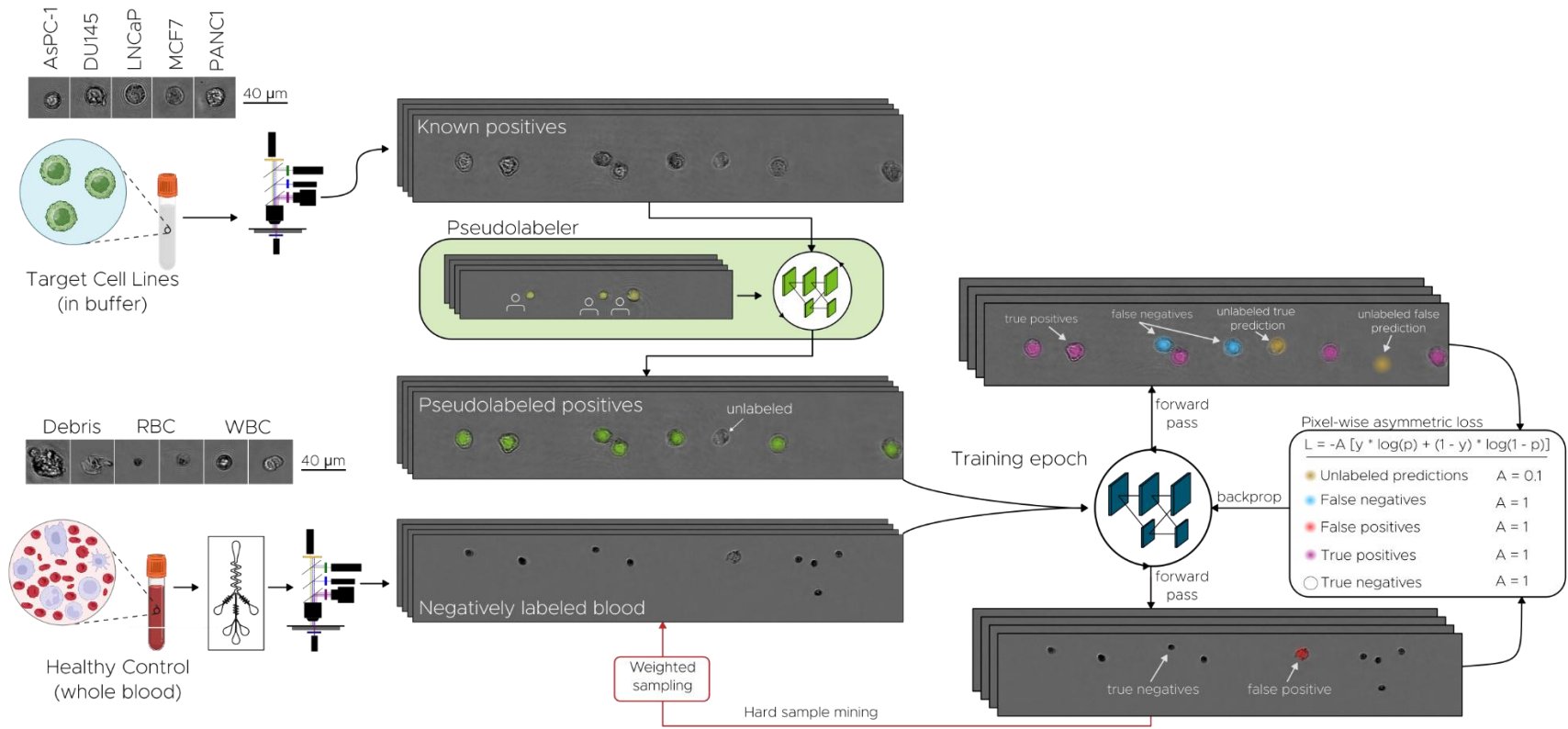

*Fig. 3 Deep learning model training process. Positive and negative samples are generated by imaging cell lines and healthy blood (post-microfluidic enrichment) in separate samples. All images derived from healthy blood are labeled negatively (i.e., lacking any Gaussian keypoint label). Cell lines are spiked into buffer, and all nucleated cells are labeled as positive via a pseudo-labeler neural network, which is itself trained from human-generated labels. The CTC detection model training. Thus although the pseudo-labeling model is trained from human-generated labels, the CTC detection model can be trained in an automated fashion, enabling large datasets. During each training epoch, the effects of false negatives from the pseudo-labeler (i.e., unlabeled cell line cells) are mitigated using pixel-wise asymmetric loss, wherein CTC model predictions lacking a corresponding pseudo-labeler label (depicted as yellow Gaussian blobs) are penalized at 1/10$^{th}$ the rate of false positives and false negatives. Each training epoch also uses hard sample mining such that images from low-performing runs from healthy controls are preferentially sampled in subsequent epochs.*

## System Validation

To evaluate the overall detection performance of the system, we conducted validation experiments using both spiked and unspiked blood samples. Each specimen was prepared using approximately 3 mL of healthy donor whole blood. One group was spiked with LNCaP prostate cancer cells (n = 18), while the control group remained unspiked (n = 18). To enable fluorescence-based confirmation, cancer cells were stained with a cell tracker dye prior to spiking, simulating an ideal condition with uniform marker expression and minimal nonspecific labeling. The unspiked control cohort was used to determine the system's false positive rate (FPR). In contrast, the spiked samples were used to calculate the recovery rate, defined here as the proportion of introduced cancer cells that were successfully identified by the system. This recovery rate reflects the true positive rate (TPR), also referred to as sensitivity. Full experimental protocols are described in SI §5.1.

To contextualize these results, we also generated a theoretical recovery curve based on a large validation dataset containing 25 cancer cell lines not used during training. As shown in Figure 4a, the recovery rate increases approximately linearly as the model confidence threshold decreases—from about 30% at a threshold of 0.9 to nearly 80% at 0.2. This trend reflects the expected trade-off between sensitivity and specificity. At higher thresholds, the system is more conservative and yields fewer detections, while lower thresholds allow more potential CTCs to be captured at the cost of admitting more false positives. Across the range from approximately 0.2 to 0.8, the experimental recovery rate was consistently lower than the theoretical prediction, with a maximum deviation of approximately 10%. This discrepancy may be attributed to biological variability between the controlled cell line conditions (i.e., cell lines in pure buffer) versus the more complex background of spiked blood samples (i.e., cell lines in blood). It also reflects necessary differences in how recovery is defined for these two different conditions (further details are available in SI §5). Nonetheless, both curves showed qualitatively similar trends, indicating that the model generalizes well across different sample contexts.

Given the expected abundance of CTCs in late-stage cancer patients, estimated at around 10 cells/mL, an operating threshold of 0.5 was chosen. At this threshold, the system achieved a TPR of around 60% while maintaining a low FPR. Lowering the threshold would improve sensitivity but introduce more false positives, which can undermine specificity in clinical use. This balance between FPR and TPR is critical for optimizing diagnostic accuracy, especially in the context of rare cell detection. Figure 4b shows the measured FPR across a range of thresholds, expressed in terms of false positive detections per mL in the control samples. Because false positive events were rare, the plot is composed of discrete points. At the chosen threshold of 0.5, only 5 false positives were detected in total across all healthy blood samples, corresponding to fewer than 0.1 false positives/mL (total unspiked sample volume was approximately 55 mL, i.e., slightly more than 3 mL per sample from the 18 samples). Even at the most permissive threshold of 0.1, the FPR remained below 1 cell/mL. Data concerning the model's FPR based on nucleated cells in healthy blood without IF data can be found in SI §3.2. This analysis helps clarify how FPR behaves in the absence of orthogonal molecular labeling and supports evaluation of system performance in strictly label-free settings.

To complete the assessment of diagnostic performance, we estimated the positive predictive value (PPV, or precision), which defines the probability that a detected cell is a true positive. While our validation experiments do not permit a direct measurement of PPV, it can be calculated from the experimentally determined TPR and FPR. For this estimation, we assumed a clinically representative CTC abundance of 10 cells/mL. At the chosen operating threshold of 0.5, this model yields a PPV of approximately 0.98. Such a high PPV underscores the system's strong specificity and indicates that detections are highly likely to be true positives, a critical feature for reliable rare cell analysis.

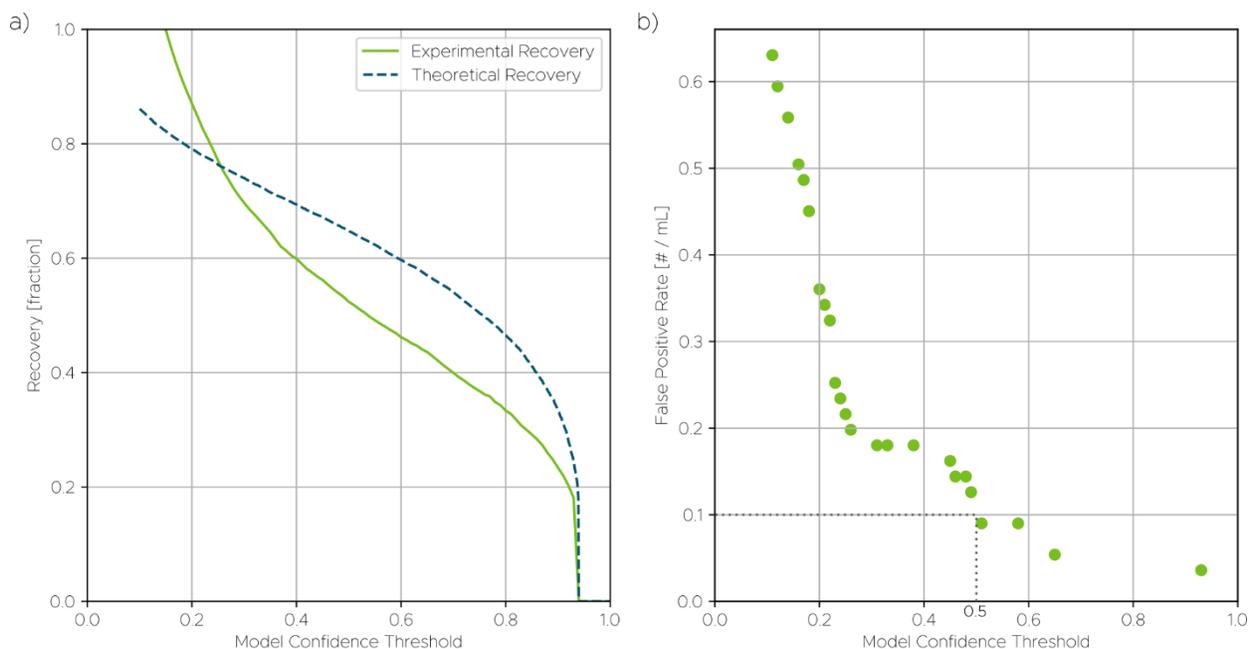

*Fig. 4 Model validation using spiked experiments. (a) Measured recovery at different operating thresholds. (b) False positive rate (FPR), shown as discrete points due to the low number of false positives detected, with only* 5 *false positive cells found across all unspiked samples (totaling 55 mL of whole blood) at the operating threshold of 0.5.*

## Patient Sample Performance

To evaluate the efficacy of our platform under clinically relevant conditions, we assessed its ability to detect CTCs in blood samples from prostate cancer patients in comparison to healthy donors. The patient cohort comprised of 13 male subjects diagnosed with metastatic castration-resistant prostate cancer (mCRPC), while the healthy control cohort consisted of 8 male donors without any known cancer diagnosis. Patient samples were subjected to staining for prostate-specific membrane antigen (PSMA) and EpCAM proteins, commonly expressed on the surface of prostate cancer cells and epithelial cells, respectively (van der Toom et al. 2019). A cell was classified as a CTC only if it exceeded the holography model threshold of 0.5 and exhibited a positive PSMA signal, as determined by PMT scoring (see SI §2.2 for details). This dual requirement was designed to ensure high tumor specificity and reduce the likelihood of misclassifying non-malignant nucleated cells. Although EpCAM expression was also measured, it was intentionally excluded from the criteria for CTC identification. This decision reflects the well-established observation that EpCAM expression can be downregulated in CTCs undergoing epithelial-to-mesenchymal transition (EMT), a process associated with increased metastatic potential and poorer prognosis (Mikolajczyk et al. 2011, Yu et al. 2013). Relying on EpCAM alone would risk missing a clinically significant subset of tumor cells. Instead, EpCAM was used as a phenotypic marker to help characterize the epithelial profile of the detected PSMA-positive cells.

This molecular gating strategy revealed a substantial proportion of EpCAM-negative CTCs: across the patient cohort, only 37% of PSMA-positive cells were also EpCAM-positive. This finding has important implications. It suggests that a large fraction of CTCs would likely evade detection in conventional EpCAM-based platforms, particularly in patients with epithelial-to-mesenchymal phenotypic shifts. By leveraging label-free holographic morphology in conjunction with PSMA, the system is capable of identifying a broader spectrum of tumor cells, including EpCAM-low or EpCAM-negative populations.

Figure 5a shows the distribution of detected CTC counts across both cohorts, with values normalized to the initial whole blood volume to account for variability in sample collection. Cancer patients exhibited a markedly higher CTC burden than healthy donors, with median counts of 12.5 cells/ml and 1.5 cells/ml, respectively (counts per patient are provided in SI §6). These findings are consistent with previously reported ranges for late-stage prostate cancer (e.g., Ried et al., 2017) and support the system's ability to reliably distinguish clinical from non-clinical samples. The spread within the patient group reflects the expected biological heterogeneity in disease progression and tumor shedding rates. Additionally, several patients exhibited CTC counts exceeding 20 cells/ml, while healthy individuals exhibited consistently low counts with few outliers, further reinforcing diagnostic discriminability. Representative holograms of CTCs detected in patient samples are shown in Figures 5b–e, along with corresponding IF signals in Figures 5f–i. The detected cells display a wide range of morphological features and antigen expression profiles, including EpCAM-negative phenotypes that are often underrepresented in traditional antigen-based CTC detection assays such as CellSearch®.

These findings suggest that the system's applicability extends to the detection of patient-derived CTCs, notwithstanding its initial training using cultured cell lines. Similar to the spiked experiments, the utilization of PSMA staining effectively decreased false positive signals in healthy samples. The results from these patient samples provide critical evidence of the system's efficacy beyond controlled lab settings and emphasizes the potential of our platform as a powerful tool for cancer diagnostics and screening.

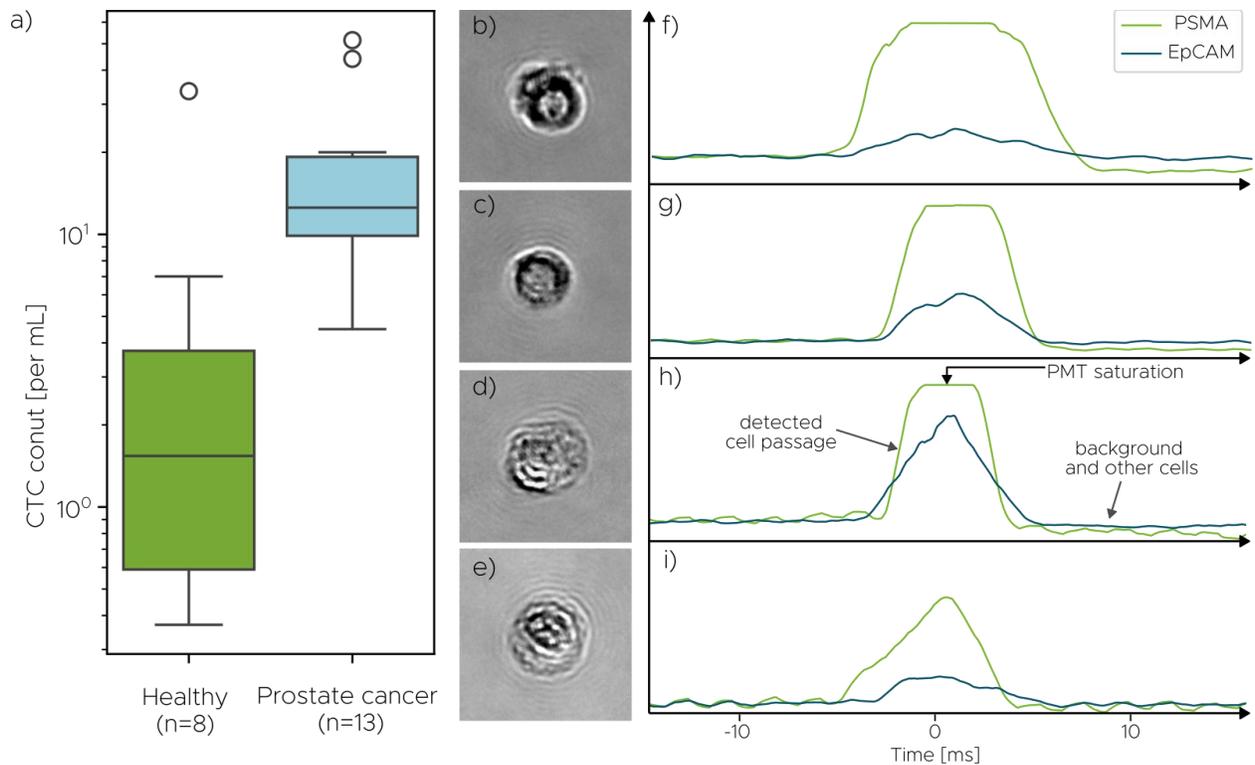

*Fig. 5 Detection rate of the model evaluated on a cohort of healthy donors and donors with late-stage prostate cancer (mCRPC). (a) Distribution of detected CTCs in patient samples, normalized to the volume of whole blood in the initial tube to account for sample variability. (b-e) CTCs identified in prostate cancer patient samples, with corresponding immunofluorescent (IF) signal (f-i). For the two IF channels captured, PSMA served as a prostate-specific marker and EpCAM served as a pan-cancer marker (for epithelial cells).*

# Discussion

## Technical Innovation and Comparison with Existing CTC Diagnostics

We present a multimodal CTC detection platform that combines inertial microfluidic enrichment, digital holographic microscopy (DHM), and immunofluorescence (IF) profiling within a deep learning–enabled framework. To our knowledge, this is the first deep learning–based DHM system to achieve both high recovery and a patient-level false positive rate below 1 cell/mL. This level of specificity is particularly critical given that CTC concentrations typically fall below 1 cell/mL in early-stage cancers and average around 10 cells/mL in advanced disease. In this context, even a modest false positive rate from the background of approximately $10^6$ WBCs/mL can produce misleading CTC signals, underscoring the importance of meeting this benchmark for any clinically deployable assay.

At the core of the platform is a customized deep learning model trained and validated on an exceptionally large and morphologically diverse dataset, comprising over 5.9 million healthy cell images and 3.1 million cancer cell images derived from 25 cell lines spanning a broad range of tissue origins. This diversity is essential for enabling robust detection of CTCs, which are not only rare but also phenotypically heterogeneous. The model's architecture supports flexible adjustment of operating thresholds, allowing tuning of sensitivity and specificity to suit different clinical scenarios. For instance, stringent thresholds can be applied in early-stage screening where false positives must be minimized, whereas more permissive settings may be appropriate for minimal residual disease monitoring or longitudinal surveillance, where detection sensitivity is prioritized.

In contrast to state-of-the-art CTC detection platforms that rely exclusively on biochemical labeling, our system leverages label-free phase-based imaging to characterize cellular morphology and optical thickness independent of surface marker expression. Antigen-based methods, such as CellSearch®, Epic Sciences platform, and other marker-driven systems, depend on the presence of antigens like EpCAM or cytokeratin, and may fail to detect CTCs that have undergone epithelial to mesenchymal or exhibit low marker expression (Mikolajczyk et al. 2001, Allard et al. 2004, Yu et al. 2013). This challenge is especially pronounced in real patient samples, where CTCs often diverge morphologically and molecularly from the cancer cell lines used for benchmarking (Powell et al. 2012, Park et al. 2014, van der Toom et al. 2016). By combining DHM with IF signal profiling, our platform captures a broader range of CTC phenotypes while preserving the specificity required for clinical deployment. The fusion of orthogonal modalities—label-free morphology and targeted molecular markers—enables robust classification even in samples with rare or atypical tumor cells.

Our system also addresses the limitations of physical property–based enrichment techniques, which rely on parameters such as size, density, and deformability to isolate CTCs from whole blood. Technologies including Parsortix®, ClearCell FX, and deterministic lateral displacement chips have shown promise for label-free enrichment, but face a fundamental trade-off in sensitivity and specificity due to overlapping physical characteristics between CTCs and WBCs (Allard et al. 2004, Chen et al. 2015, Smit and Pantel 2024). As with biochemical markers, the morphological and mechanical properties of CTCs in patients can differ significantly from those of cultured cell lines (Park et al. 2014, van der Toom et al. 2016). In our approach, inertial microfluidics is used primarily to reduce sample complexity by removing RBCs and reducing WBC density. However, final classification is not determined by mechanical proxies like size or deformability. Instead, we use image-based deep learning trained on millions of annotated cells. This enables high-fidelity detection of phenotypically diverse CTCs, including those that would be missed by purely physical or biochemical filters.

A key advantage of our platform is its scalability and compatibility with clinical workflows. The entire process—from whole blood collection to quantitative CTC enumeration—is completed within six hours per 10 mL tube, without requiring manual cell inspection or operator-dependent interpretation. The system remains robust when processing samples up to 24 hours post-collection, even when using EDTA tubes, mitigating challenges posed by pre-analytical variability in blood quality. This is in contrast to widely used systems such as Parsortix®, which typically require same-day processing when using EDTA tubes due to sensitivity to clotting or cell degradation, thereby limiting operational flexibility in real-world settings (Qin et al. 2014, Payne et al. 2021). In addition, our compact hardware footprint, low reagent consumption, and minimal technician oversight enable high-throughput parallelization across multiple stations. These features collectively position the system for scalable deployment in clinical laboratories, including large-cohort studies and longitudinal monitoring applications.

## Toward Broad Clinical Utility in Liquid Biopsy

Our platform is inherently adaptable to accommodate a wide range of clinical workflows and biological sample types. The diversity of cancer cell lines included in the training dataset supports the model's potential to detect CTCs originating from tumor sites beyond the prostate. Antigens can be selectively incorporated to target cells from specific cancer types or to guide treatment decisions, such as assessing ER/PR or HER2 status in breast cancer. Holographic imaging, being label-free and non-destructive, is compatible with any optically transparent medium. Although this study focused on CTC detection in peripheral blood, the same analytical pipeline can be seamlessly extended to other clinically relevant biofluids such as saliva, urine, pleural effusion, and cerebrospinal fluid. This versatility offers a promising foundation for developing noninvasive cancer diagnostics applicable across diverse tumor types and anatomical compartments (Patel et al. 2011, Yang et al. 2014, Andersson et al. 2014, Gostomczyk et al. 2024). By integrating label-free DHM with targeted IF, the platform enables robust detection of rare tumor cells, independent of sample origin or matrix composition.

Collectively, these features enable the system to transform a complex, billion-cell blood sample into a concise, high-confidence readout—without requiring expert interpretation by trained technicians or pathologists. A key advantage of the workflow is its real-time detection capability, which operates during continuous sample flow. This live imaging approach not only accelerates CTC enumeration but also preserves cell viability, thereby enabling seamless integration with subsequent isolation and phenotypic or molecular characterization.

This operational flexibility positions the platform for broad utility across the cancer care continuum, including early detection, treatment monitoring, and post-therapy surveillance. The ability to process standard clinical blood volumes with high specificity and rapid turnaround time is particularly advantageous for longitudinal tracking of minimal residual disease and early relapse. Beyond enumeration, the platform enables recovery of viable CTCs for single-cell analyses, including genomic, transcriptomic, and proteomic profiling. These capabilities are well aligned with emerging integrative workflows such as VERSA (Versatile Ex Vivo Rare-cell Sorting and Analysis), which aim to dissect tumor heterogeneity, assess drug sensitivity ex vivo, and support personalized therapy design (Lang et al. 2017, Pezzi et al. 2018).

Looking forward, a key direction for future development is the realization of a fully label-free diagnostic workflow powered entirely by deep learning. As the training dataset grows to encompass a broader range of patient-derived CTC morphologies, we anticipate model performance will reach the threshold needed to eliminate dependency on IF labeling. A transition to a purely image-based pipeline would significantly streamline sample processing, reduce reagent use, and shorten assay turnaround, thereby enhancing feasibility in decentralized, resource-limited, or point-of-care settings (Ring et al. 2023, Smit and Pantel 2024). Moreover, such a system could enable pan-cancer detection and even subtype classification, relying exclusively on morphological and phase-based features. This shift from antigen-dependent identification to phenotype-aware optical profiling would represent a fundamental advance in the capabilities of liquid biopsy technologies.

# Methods

## Sample Collection and Preparation

Blood samples were collected into EDTA tubes to prevent clotting. Each donor provided at least two 10 mL tubes of whole blood which were then processed with a patented passive RBC depletion and CTC enrichment method, reducing the sample to 3 mL per tube (patent no. WO2024064911A1; further description in SI §1). The enriched samples were stained for protein markers PSMA and EpCAM following an IF staining protocol. Other proprietary steps were performed that are not pertinent to this study. All samples are completely processed the day following collection. The active processing time of each sample is approximately 6 hours.

Artificial CTC-laden blood samples, used in recovery experiments, were created by spiking a targeted number of LNCaP cells into a healthy blood sample following the microfluidic enrichment procedure. The true number of cells observed by the station was validated by collecting the sample from the outlet (i.e., after DHM imaging) into a well plate and counting on a fluorescence microscope. For more details on the spiking procedure and analysis, see SI §5.1 and §5.2.

Experimental protocols were approved by the Institutional Review Board (IRB) of the University of Minnesota. The study methods were carried out in accordance with all relevant guidelines and regulations. All patients and donors included in this study provided informed consent prior to having their blood drawn.

## Holographic Imaging Station

Digital holographic microscopy was performed using a Gabor holography setup that did not separate the object and reference waves (Gabor 1972, Xu et al. 2001). The light source was a pulsed laser diode (405 nm wavelength), and images were obtained using 10x infinity corrected microscope objective with 0.30 numerical aperture (NA). The focal plane was adjusted to be aligned with the depth-wise center of the flow channel, and a focus quality metric was computed and provided to the user in real-time to help guide this process (further details in SI §2.2). Holograms were captured with a 1.6 mega-pixel machine vision camera with a pixel pitch of 3.45 µm. The camera was synchronized with the laser (450 frames-per-second) with a capture region of interest (ROI) of 1440×256 pixels. Images were streamed over USB to the capture and processing computer. There were 18 imaging stations used for the study to avoid over-fitting to station-specific attributes such as the beam profile or dust on the optics. For further details of the system, see SI §7.

# Deep Learning Model Development

CTCs are identified in holographic images using a trained deep learning model. Here we outline the architecture and training of the model, including strategies for annotation, sample selection, and postprocessing. We describe the model's design adaptations for high-throughput inference, the use of pseudo-labeling for large-scale training without extensive manual annotation, and the incorporation of hard sample mining to address class imbalance. Details on the training and validation datasets, including image acquisition and quality control procedures, are also provided.

## Model Architecture and Hyper Parameters

We optimized the HRNet architecture (Wang et al., 2020) to support real-time inference while preserving high classification accuracy. This involved tailoring the number of layers to balance computational efficiency and feature richness. Specifically, the coarsest layers were pruned to reduce the receptive field, ensuring that the model focused on spatial features most relevant to small target cells. This adjustment allows the model to extract sufficient discriminative features while minimizing the influence of surrounding objects. The model weights were initialized using public weights pre-trained on the COCO dataset. The Adam optimizer was used with a learning rate of 0.01 and exponential learning rate decay by a factor of 0.85 each epoch.

The pixel-wise loss function is an asymmetric binary cross-entropy loss defined as

$$L = -A[y * log\,(p) + (1-y) * log\,(1-p)\,]$$

where A=0.1 for false positives on positive samples (cell lines) and A=1 otherwise.

## Training Image Annotation

Each image was pre-processed by subtracting a background reference, which was computed using an exponential moving average of preceding frames within the same imaging session. Following background subtraction, the image was normalized using a fixed mean and standard deviation to ensure consistent intensity scaling. Target heatmaps were generated using a pre-trained pseudo-labeling model. For each detected cell location, a Gaussian blob with a standard deviation of 2.8 µm was placed to define the spatial confidence region. This representation enabled the binary cross-entropy loss function to emphasize not only the presence of a detection but also allow precise localization accuracy.

## Hard Sample Mining

The dataset was inherently imbalanced: many images contained no cells, and healthy blood samples were often dominated by RBCs, which are generally easier to distinguish from CTCs than WBCs. To address this challenge, a hard sample mining strategy was applied. An earlier version of the model, trained on a smaller but similar dataset, was used to pre-screen the training images. Only those images with at least a moderate likelihood of containing a true positive detection, based on a confidence threshold of 0.3, were retained. This step significantly reduced the size of the training set while enriching it with more informative and ambiguous examples. During training, batches were drawn preferentially from imaging sessions that showed poor model performance in the previous epoch, with sampling weighted by the average loss. This prioritization ensured that the most difficult examples were repeatedly encountered during training. To manage computational efficiency, performance evaluation was conducted at the imaging session level instead of re-scoring individual images after each epoch.

### Training Dataset

The training dataset was composed of healthy blood samples collected from 32 unique donors, some of whom contributed multiple samples. Positive samples were generated using five established cancer cell lines: AsPC-1 and PanC1 (pancreatic), DU 145 and LNCaP (prostate), and MCF7 (breast) (SI §4.2 Table 2). In total, the dataset was compiled from over 140 imaging sessions, spanning more than 200 hours of acquisition and yielding approximately 300 million raw images. As many of these images were either empty or trivial to classify, an earlier version of the detection model was used to pre-filter the dataset. Images with a model confidence score above 0.3 were retained, resulting in a curated dataset of 7.9 million images, including 5.9 million from healthy samples and 2.0 million from cancer cell lines. These were divided into training and validation sets using an 80:20 split. To prevent data leakage, all images from a given acquisition session were assigned to the same partition, ensuring that repeated captures of the same cell did not appear in both sets. Multiple imaging stations were used during data collection to introduce variability in optical and hardware conditions, thereby improving model generalizability. Importantly, the only manually annotated data used for training came from a small pseudo-labeling dataset consisting of 2500 images. All other annotations in the training dataset were generated automatically.

### Validation Dataset

The spiked validation dataset consisted of 36 samples, yielding a total of 185 million images. In addition, 1.1 million images containing isolated cancer cell lines (i.e., spiked into buffer, not blood) were acquired separately to evaluate the model's theoretical recovery in the absence of background interference or blood-related losses. True positive detections were defined by spatial proximity between the prediction peaks generated by the detection model and those produced by the pseudo-labeling model. This comparison enabled a quantitative estimate of the model's theoretical recovery, expressed as the true positive rate (TPR), in an idealized context without confounding biological noise (Fig. 4).

# Data Availability

The data that support the findings of this study are available from the corresponding author upon reasonable request.

# Code Availability

The code used for data acquisition and analysis is available from the corresponding author upon reasonable request.

# Bibliography

epitope-agnostic isolation of circulating tumor cells. *Proceedings of the National Academy of Sciences*, 117(29), 16839-16847.

33. Nagrath, S., Sequist, L. V., Maheswaran, S., Bell, D. W., Irimia, D., Ulkus, L., ... & Toner, M. (2007). Isolation of rare circulating tumour cells in cancer patients by microchip technology. *Nature*, 450(7173), 1235-1239.
34. Nasiri, R., Shamloo, A., & Akbari, J. (2022). Design of two Inertial-based microfluidic devices for cancer cell separation from blood: A serpentine inertial device and an integrated inertial and magnetophoretic device. *Chemical Engineering Science*, 252, 117283.
35. Okano, H., Konishi, T., Suzuki, T., Suzuki, T., Ariyasu, S., Aoki, S., ... & Hayase, M. (2015). Enrichment of circulating tumor cells in tumor-bearing mouse blood by a deterministic lateral displacement microfluidic device. *Biomedical Microdevices*, 17, 1-11.
36. Ozkumur, E., Shah, A. M., Ciciliano, J. C., Emmink, B. L., Miyamoto, D. T., Brachtel, E., ... & Toner, M. (2013). Inertial focusing for tumor antigen–dependent and–independent sorting of rare circulating tumor cells. *Science Translational Medicine*, *5*(179), 179ra47.
37. Pantel, K., & Alix-Panabières, C. (2010). Circulating tumour cells in cancer patients: challenges and perspectives. *Trends in Molecular Medicine*, 16(9), 398-406.
38. Pantel, K., & Alix-Panabières, C. (2019). Liquid biopsy and minimal residual disease—latest advances and implications for cure. *Nature Reviews Clinical Oncology*, 16(7), 409-424.
39. Park, S., Ang, R. R., Duffy, S. P., Bazov, J., Chi, K. N., Black, P. C., & Ma, H. (2014). Morphological differences between circulating tumor cells from prostate cancer patients and cultured prostate cancer cells. *PloS ONE*, 9(1), e85264.
40. Patel, A. S., et al. (2011). Identification and enumeration of circulating tumor cells in the cerebrospinal fluid of breast cancer patients with central nervous system metastases. *Oncotarget*, 2(10), 752.
41. Payne, Karl, et al. (2021). Immediate sample fixation increases circulating tumour cell (CTC) capture and preserves phenotype in head and neck squamous cell carcinoma: towards a standardised approach to microfluidic CTC biomarker discovery. *Cancers*, 13(21), 5519.
42. Pezzi, Hannah M., et al. (2018). Versatile exclusion-based sample preparation platform for integrated rare cell isolation and analyte extraction. *Lab on a Chip*, 18(22), 3446-3458.
43. Pirone, D., Montella, A., Sirico, D. G., Mugnano, M., Villone, M. M., Bianco, V., ... & Ferraro, P. (2023). Label-free liquid biopsy through the identification of tumor cells by machine learning-powered tomographic phase imaging flow cytometry. *Scientific Reports*, 13(1), 6042.
44. Powell, A. A., et al. (2012). Single cell profiling of Circulating tumor cells: Transcriptional heterogeneity and diversity from breast cancer cell lines. *PLoS ONE*, 7(5), e33788.
45. Qin, J., Alt, J. R., Hunsley, B. A., Williams, T. L., & Fernando, M. R. (2014). Stabilization of circulating tumor cells in blood using a collection device with a preservative reagent. *Cancer Cell International*, 14, 1-6.
46. Ring, A., Nguyen-Sträuli, B. D., Wicki, A., & Aceto, N. (2023). Biology, vulnerabilities and clinical applications of circulating tumour cells. *Nature Reviews Cancer*, 23(2), 95-111.
20

# Acknowledgements

Research reported in this publication was supported by the National Cancer Institute of the National Institutes of Health under Award Number R41CA268344. The content is solely the responsibility of the authors and does not necessarily represent the official views of the National Institutes of Health.

Portions of this work were conducted in the Minnesota Nano Center, which is supported by the National Science Foundation through the National Nanotechnology Coordinated Infrastructure (NNCI) under Award Number ECCS-2025124.

E.S.A. is partially supported by NCI grant P30 CA077598 and DOD grant W81XWH-22-2-0025.

The authors gratefully acknowledge the contributions of Jayant Parthasarathy and the broader team at Astrin Biosciences, including Alan Avilez-Olvera, Sullivan Bluhm, Joshua Dean, Adam Groth, Alec Horrmann, Erik Keohane, Madelynn Miller, Kara Paulsen, Carissa Rungkittikhun, James Uden, and Corina Valencia.

# Author Contributions

Kevin Mallery: writing, data analysis

Nathaniel R. Bristow: writing, data analysis, imaging device development

Nick Heller: machine learning model development, software development

Yash Travadi: data analysis, machine learning model development

Ali Arafa: sample procurement

Kaylee Kamalanathan: data collection, protocol development, microfluidic device development

Catalina Galeano-Garces: data collection, protocol development

Mahdi Ahmadi: microfluidic device development

Grant Schaap: data collection

Alexa Hesch: data collection

Olivia Hedeen: data collection

Zikora Izuora: data collection

Joel Hapke: protocol development

Jeffrey Miller: software development

Arjun Viswanathan: machine learning model development

Ivo Babris: software and device testing

Songyi Bae: data collection

Tuan Le: software architecture

Tony Clacko: operations management

Emmanuel S. Antonarakis: sample procurement

Badrinath R. Konety: operations management

Justin M. Drake: conception and design, study supervision, laboratory operations management

Jiarong Hong: conception and design, writing, data analysis, engineering operations management, imaging device development, machine learning model development

# Supplemental Information (SI)

## 1 Inertial Enrichment Chip

### 1.1 Intent & Purpose

The inertial enrichment step is necessary to reduce the overall number of cells seen by the system. The undiluted cell concentration in whole blood is extremely high – approximately $10^6$ WBCs/mL and $10^9$ RBCs/mL. The system was designed to work best when there is approximately 1 cell per hologram (thus ensuring an unambiguous fluorescence signal). An undiluted sample has thousands of cells in the 2.5 nL ROI, necessitating some form of dilution. This population is also dominated by RBCs which, while unlikely to be confused with CTCs, still contaminate the signal. Inertial enrichment was able to not only deplete RBCs, but to concentrate CTCs while retaining many of the WBCs. The enrichment ensured that CTCs formed a larger fraction of the sample, further reducing the impact of false positives.

### 1.2 Mechanism of Separation

As its name suggests, inertial enrichment separates cells primarily based on their mass, allowing for the removal of the much smaller RBCs. It also separates based on other mechanical properties such as deformability which enables the specialized concentration of CTCs.

Flow through curved channels can develop secondary motion in the form of counter-rotating vortices, known as Dean vortices. This has been observed to lead to the separation and even trapping of particles based on density. Repeated bends in a channel ultimately lead to spatial separation of the cells, enabling mechanical separation. For a full review on the topic, see Zhang et al. (2016).

### 1.3 Design Specifications

Inertial enrichment chips were fabricated from PDMS bonded to glass microscope slides. For further specifications on channel geometry, see patent (WO2024064911A1).

### 1.4 Cell Line Recovery

The retention of cancer cells processed through the inertial chip was validated with a mass-balanced recovery experiment. Approximately 3000 A549 cells were spiked into buffer media and run through the inertial device. The collection and waste outlets were collected and distributed to a well plate for counting. The experiment was repeated for 6 chips, each tested in triplicate. The recovery counts are shown in SI Table 1.

The median recovery is 95%. There were two replicates that failed with unusually low recovery. Generally, these failures are due to clogs or foreign material blocking the flow channel. When processing blood, these instances are readily identifiable by a visible change in the color of the collected sample due to an increase in the red blood cell concentration.

*SI Table 1. Results of the inertial chip mass balance recovery experiments.*

| Chip ID | Replicate | Waste Count | Collect Count | Total Count | Recovery % |
|---|---|---|---|---|---|
| 1 | 1 | 160 | 2704 | 2864 | 94% |
| 1 | 2 | 143 | 3098 | 3241 | 96% |
| 1 | 3 | 83 | 3029 | 3112 | 97% |
| 2 | 1 | 90 | 3004 | 3094 | 97% |
| 2 | 2 | 3320 | 14 | 3334 | 0% |
| 2 | 3 | 94 | 3016 | 3110 | 97% |
| 3 | 1 | 97 | 2877 | 2974 | 97% |
| 3 | 2 | 155 | 3043 | 3198 | 95% |
| 3 | 3 | 228 | 4621 | 4849 | 95% |
| 4 | 1 | 83 | 2808 | 2891 | 97% |
| 4 | 2 | 2975 | 52 | 3027 | 2% |
| 4 | 3 | 51 | 3072 | 3123 | 98% |
| 5 | 1 | 74 | 2806 | 2880 | 97% |
| 5 | 2 | 299 | 2660 | 2959 | 90% |
| 5 | 3 | 140 | 5535 | 5675 | 98% |
| 6 | 1 | 187 | 2437 | 2624 | 93% |
| 6 | 2 | 28 | 143 | 171 | 84% |
| 6 | 3 | 213 | 3050 | 3263 | 93% |

# 2 Data processing

## 2.1 PMT Signal

Raw fluorescence intensity signal captured by the photomultiplier tubes (PMTs) was preprocessed prior to peak detection. First, the timestamps associated with the PMT data were synchronized to the camera timestamps via periodic, controlled synchronization "events" that were readily detected in the PMT signal and the digital holographic microscopy (DHM) camera signal. Thus, any latency between the PMT and DHM data could be adjusted for when matching holographic model detections to spikes in fluorescence, to within a sub-frame capture timescale.

Fluorescent events were detected by first removing the background signal followed by convolution with a template signal, providing a "matched filter" for the target detection timescale. Peaks in the signal could then cleanly be detected via standard peak-finding algorithm based on prominence and width. PMT scoring provided an arbitrary but consistent scale for comparing detections against each other for fluorescence intensity. For patient samples, a PMT spike had to have a score > 8 to be identified as a PSMA positive cell.

## 2.2 Quality Control

Although the Detection Model was trained and augmented using a variety of image quality conditions, including variability in optical focus quality for cells, an automated method for assessing data quality was developed to ensure consistent results. Quantifying focus quality in holograms is notoriously difficult due to interference fringes that make standard metrics (e.g., sharpness) insufficiently robust. Instead, the cell focus was assessed indirectly using the holograms produced by small (sub-micron) debris particles that end up distributed randomly in the microfluidic channel. Debris on either side of the focal plane have a clearly distinct appearance and should be present in equal numbers around cells that are in focus. Using the same HRNet-based architecture as the Detection Model, a model for detecting such debris particles was developed from hand-labeled datasets to provide focus quality metric on a frame-by-frame basis. This focus quality metric was computed live on a separate server and made visible to the laboratory technician during data acquisition to ensure data quality consistency.

# 3 Deep Learning Model

## 3.1 Architecture

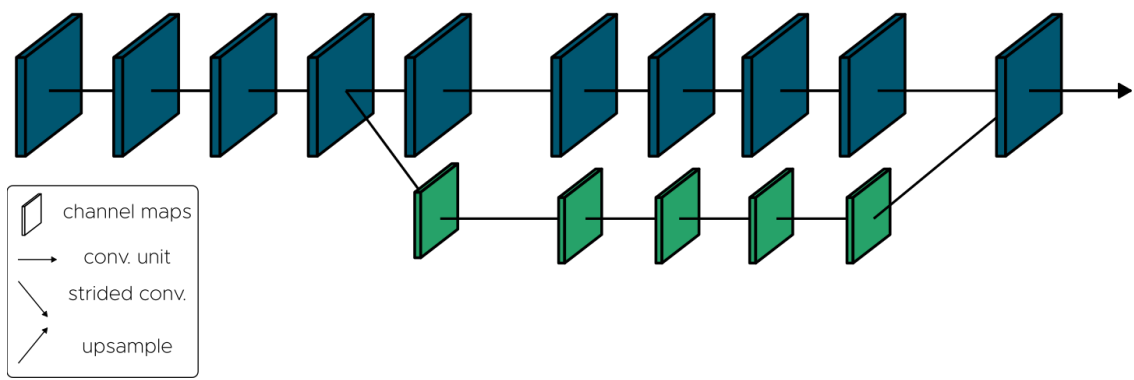

*SI Fig. 1 Convolutional neural network used for real-time in-flow CTC detection.*

The convolutional neural network (CNN) used by the DHM detection model was based on the original HRNet architecture proposed by Wang et al. (2020). To achieve the desired image processing throughput, the network was customized to the form shown in SI Fig. 1, after removing two levels of strided convolutions. The pixel-wise confidence heatmap output by this CNN was then used to detect cells of interest by applying a peak detection algorithm to yield confidence and position for each cell candidate.

## 3.2 Model validation

To supplement the System Validation results presented in the main text, we here present the false positive rate without the requirement that detections be IF positive. SI Fig. 2 corresponds to the same experimental data presented in Fig. 4b, where a false positive is any detection seen in the healthy samples.

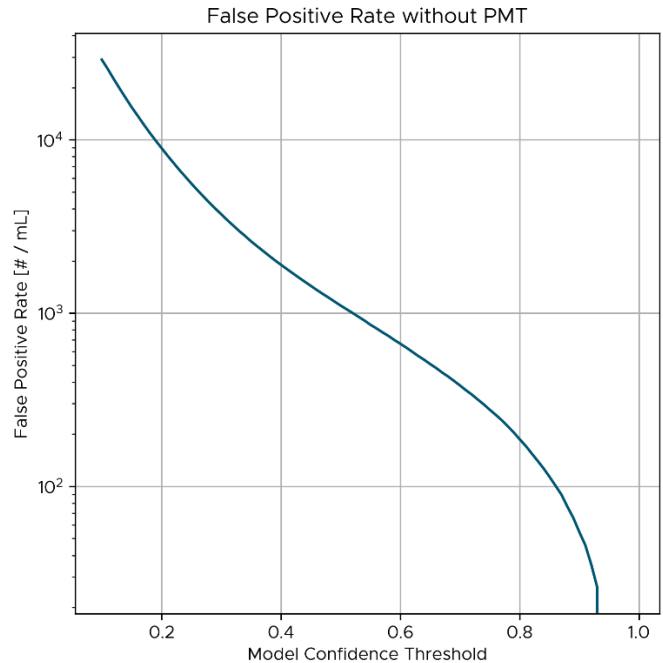

*SI Fig. 2 DHM detection model (i.e., without integration of immunofluorescence signal captured with PMT) false positives for varying model confidence operating points from recovery experiment control samples.*

# 4 Cell Lines and Blood Sample Examples

## 4.1 Image Examples

This section provides representative examples of both negative (healthy) and positive (cancer) cells, along with other objects encountered during model training. These examples illustrate the diverse morphological characteristics of CTCs and other cellular components found in blood samples.

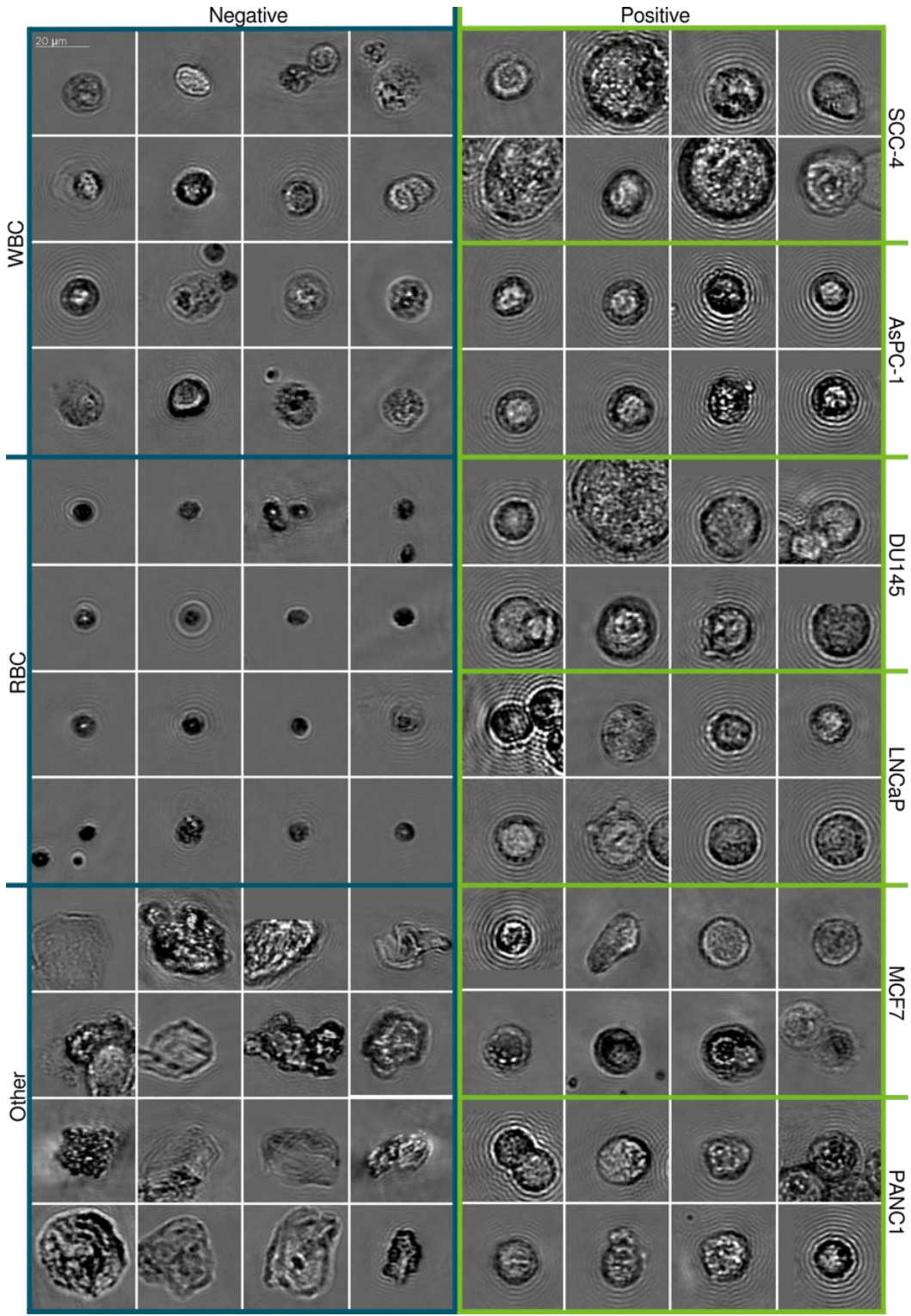

SI Fig. 3 Sample of negative (healthy) and positive (cancer) cells and objects seen during model development. Cell lines listed do not constitute an exhaustive list of all those used.

## 4.2 Cell Lines Used

*SI Table 2. Cell lines used in model development and recovery data from Fig. 4a. Note that additional cell lines were used in theoretical recovery dataset, listed in SI Table 3.*

| Name | Cancer Tissue | Training | Model Validation | Experimental Recovery (Fig. 4a) | Theoretical Recovery (Fig. 4a) |
|---|---|---|---|---|---|
| AsPC-1 | pancreas | X | X | | X |
| DU 145 | prostate | X | X | | X |
| LNCaP | prostate | X | X | X | X |
| MCF7 | breast | X | X | | X |
| PANC1 | pancreas | X | X | | X |
| SCC4 | oral | | X | | X |

*SI Table 3. Additional cell lines used in theoretical recovery (Fig. 4a)*

| Name | Cancer Tissue |
|---|---|
| A-375 | melanoma |
| A-498 | kidney |
| A-549 | lung |
| AU-565 | breast |
| BT-20 | breast |
| BxPC-3 | pancreas |
| CA-SK1 | cervical |
| H-69 | lung |
| HepG2 | liver |
| Hs 578T | breast |
| MIA PaCa-2 | pancreas |
| PC-3 | prostate |
| SCC-01 | oral |
| SCC-9 | oral |
| SCC-15 | oral |
| SCC-25 | oral |
| SNU-16 | gastric |
| T-24 | bladder |
| T-84 | colorectal |
| ZR-75-1 | breast |

Cell lines were chosen from a variety of cancers in order to maximize the diversity of cancer cell morphology that the model is trained on. True CTCs are rare, and therefore it is infeasible to train a supervised model on true CTC images. Thus, given the model's reliance on cell lines as a proxy, it was important for the model not to overfit to one particular cell line, as individual cell lines may not exactly match the morphological heterogeneity present in CTC populations. Therefore, multiple cancers are represented, with multiple cell lines for prostate cancer in particular, to hedge against this uncertainty.

# 5 Recovery Experiments

This section describes the methodology used to validate detection performance through controlled spiking experiments.

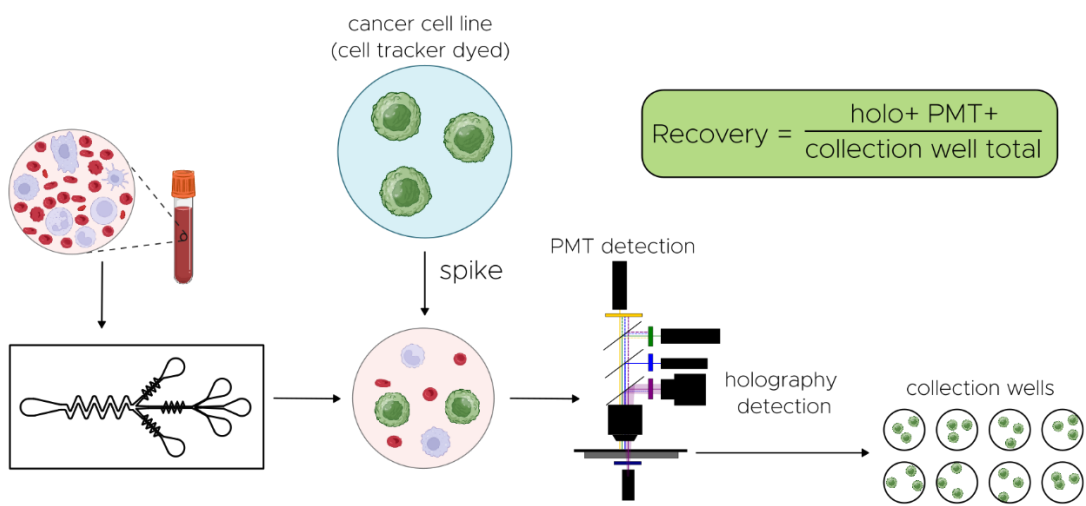

*SI Fig 4. Spiking overview for recovery experiments for model performance validation.*

## 5.1 Preparation of Spiked Samples

LNCaP cells were cultured in the same manner as those used for model training. These cells were stained with cell tracker green dye, prior to spiking into an enriched blood sample (i.e., after microfluidic depletion of RBCs and some WBCs). The cell tracker dye allowed for easy identification of spiked cells in the sample collected after imaging. Spiked cell numbers were targeted using a multi-step dilution procedure and confirmed by checking additional cell dilutions in a well plate counted under a fluorescent microscope.

## 5.2 Sample Processing & Analysis

Spiked blood samples were accompanied by control samples that did not have LNCaP cells spiked, from the same starting blood specimen. All samples, regardless of spiking, were pumped through a straight-channel microfluidic chip at the same flow rates as used for patient samples. The outlet volume of the sample was then transferred to a series of well plates and counted via a standard thresholding algorithm based on cell tracker dye expression. This method accounted for any discrepancies between the targeted spike number and the true spiked number, and enabled accurate measurement of the true positives seen by the imaging system. Thus, the experimental recovery reflected the true model performance on this cell line.

Data collected via the PMTs were evaluated only on channel 1 (FITC equivalent), as there was only one stain used. Thus, PMT 2 was ignored as some of the cell tracker dye emissions bled into the second channel and PMT 2 was not expected to be negative.

## 5.3 Theoretical Recovery Analysis

The theoretical recovery was computed using a sample of 2 million cell line images collected during model training. Twenty additional cell lines were chosen but were not included in the training data set (see SI Table 3). Unlike the experimental spiked recovery, the theoretical recovery is estimated using the pseudo-labelled counts as the denominator. This difference in definition may account for the approximately 10% difference in the experimental and theoretical recovery curves (Fig. 4).

# 6 Patient and Healthy Sample Counts

This section provides a breakdown of the sample sizes, patient demographics, and key findings.

## 6.1 Cohort Characteristics

*SI Table 4. Patient cohorts for blood samples used across model development and testing.*

| Cohort | No. of Individuals | Cancer Type | Gender | Source |
| --- | --- | --- | --- | --- |
| Training Healthy | 18 | None | Male and female | Drawn from internal volunteers and purchased from ZenBio |
| Spike Validation | 6 | None | Male | Drawn from internal volunteers |
| Test Healthy | 8 | None | Male | Drawn from internal volunteers (n=2), Purchased from ZenBio (n=6) |
| Test Cancer | 13 | Prostate (Stage 4) | Male | Collected by collaborators at University of Minnesota |

## 6.2 Per Patient CTC Enumeration Results

*SI Table 5. Healthy patient and prostate cancer patient CTC counts.*

| Patient ID | Cancer Type | CTC count per mL |
|---|---|---|
| H01 | Healthy | 7.01 |
| H02 | Healthy | 33.51 |
| H03 | Healthy | 0.37 |
| H04 | Healthy | 2.13 |
| H05 | Healthy | 2.65 |
| H06 | Healthy | 0.95 |
| H07 | Healthy | 0.61 |
| H08 | Healthy | 0.53 |
| P01 | Prostate | 11.6 |
| P02 | Prostate | 18.59 |
| P03 | Prostate | 8.36 |
| P04 | Prostate | 12.36 |
| P05 | Prostate | 20 |
| P06 | Prostate | 44.02 |
| P07 | Prostate | 9.88 |
| P08 | Prostate | 12.53 |
| P09 | Prostate | 51.53 |
| P10 | Prostate | 13.96 |
| P12 | Prostate | 4.49 |
| P12 | Prostate | 19.24 |
| P13 | Prostate | 5.11 |

# 7 Station Description & Imaging Setup

## 7.1 System

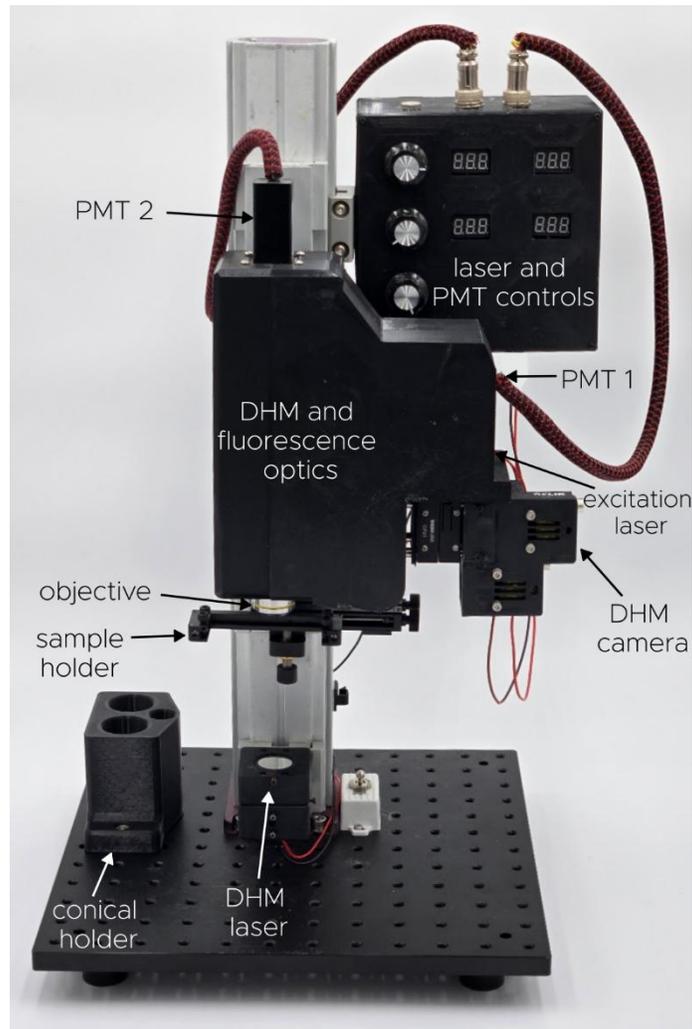

*SI Fig. 5 Dual-modality imaging system, combining digital holographic microscopy (DHM) and fluorescence sensing with two photomultiplier tubes (PMTs).*

## 7.2 Coherence & Stability Considerations

As a coherent imaging technique, DHM is highly sensitive to the quality of the illumination beam. The coherence length plays an important role – a more coherent source will improve the effective depth of focus of the system. This makes the station less sensitive to variations in the sample placement at a cost of increased noise from defects in optical components such as the dichroic mirrors and objective. Often, these defects can be removed by subtracting out a time-averaged background image. This enhancement further adds a requirement for high optical stability to ensure that the background is constant during the averaging period (approximately one second). The laser for this study was selected through a process of trial and error where we qualitatively tested the image quality of several candidate lasers.

## 7.3 Fluorescence

Fluorescent excitation was induced using a single continuous laser diode with 488 nm center wavelength and 10 nm bandwidth. The laser beam was directed into the microscope objective (and then onto the sample) using a dichroic mirror with cut-on wavelength of 510 nm. This focused the laser onto the sample with a spot size measuring approximately 0.5 mm in diameter to induce fluorescent emission from labeled cells.

Photomultipliers tubes (PMTs) from Hamamatsu (H10722-20) were used to capture fluorescent emissions in a frequency bandwidth from DC to 20 kHz. Analog voltage signals produced by the low-noise amplifiers integrated into each PMT were captured on the same machine as the image acquisition via an analogue-to-digital converter (ADC). Each PMT captured emissions in one of two wavelengths: 525 nm, corresponding to the commonly used fluorescein isothiocyanate (FITC) channel, and 575 nm corresponding to the phycoerythrin (PE) channel. In each case, a combination of dichroic mirrors and bandpass optical filters were used to separate emissions reaching each PMT and to filter out light from the excitation laser and DHM laser. For the first PMT, capturing the FITC channel, a long-pass dichroic mirror with cut-on wavelength of 552 nm reflected shorter wavelength emissions through a 525 nm center wavelength bandpass filter with 25 nm full width-half max. For the second PMT, capturing the PE channel, longer wavelengths transmitted through the same 552 nm long-pass dichroic mirror were filtered through a 575 nm center wavelength bandpass filter with 25 nm full width-half max.

## 7.4 Microfluidics & Flow System

### 7.4.1 Chip Interface with Imaging System

Imaging was performed on a PDMS microfluidic chip similar to that used during microfluidic inertial enrichment (SI §1). For this application, sample was pumped along a straight channel and a sheath flow was also used to ensure all cells are concentrated in the center of the channel, away from channel walls which might otherwise produce image artifacts. Fluid was pumped through the channel using programmable syringe pumps (Pump Systems Inc., NE-500). The sample flow rate was 6 µL/min while the sheath flow rate was 7.5 µL/min.

During sample processing the microfluidic chip was mounted in a standard microscope slide holder (Thorlabs XYF1) allowing field-of-view (FOV) adjustments in the x-y plane, while focusing adjustment was enabled by a separate linear stage (Optics Focus MAX-B34C-13S) aligned with the z-axis. Alignment of the chip to the target FOV was performed manually using markers on the chip.

### 7.4.2 Photobleaching

The PDMS used for the microfluidic chip exhibited autofluorescence which may have adversely affected the SNR of true cell fluorescence peaks. To reduce this, we photobleached all chips for at least 1.5 hours prior to processing a sample. The bleaching effectiveness was confirmed by measuring the decay of the autofluorescent signal (via the PMTs) over time. The chip was considered to be fully photobleached once the signal reached an asymptotic minimum.

## 7.5 System Footprint

The full system consisted of 3 connected components: the optical assembly (primary station), sample pumps, and computer. The optical assembly measured 30 cm x 30 cm x 60 cm (Width x Depth x Height) and sat on a lab bench. The sample syringe pumps were mounted on a shelf above the station while the acquisition computer (a standard desktop case) sat underneath the bench.

## 7.6 Automation & Control Software

A purpose-built software application (*HAIstack*) was built to operate the stations. This software allows the operator to document sample properties, view the live camera feed, control the pumps, and start and stop acquisition. This same interface also serves as the hub for the laboratory inventory management system (LIMS), including protocols, specimen database entries, data acquisition, session scheduling, etc.

The most labor-intensive aspects of the developing system were the assembly of the microfluidic chip (including inserting the sample, outlet, and waste tubes), alignment of the chip to the camera FOV, and monitoring for rare instances of leaks or other failures due to manufacturing defects. These processes can all be readily automated using robotics or through the design of specialized assembly hardware. Furthermore, the throughput of the system could be increased through the use of higher flow rates and corresponding imaging frame rates. The current system is bottlenecked by inference speeds at runtime, as only a subset of the acquired frames is saved, and used in later postprocessing, based on model confidence. Separating acquisition and inference is costlier in terms of the data pipeline required for storing images of the entire fluid volume, frame by frame, but would enable faster flow rates.

# 8 Supplementary References